\documentclass[twocolumn,pre,amsmath,amssymb,showpacs,aps]{revtex4-1}

\usepackage{graphicx}
\usepackage{color}
\usepackage{epstopdf}
\usepackage{algorithmic}

\begin{document}

\preprint{APS/123-QED}

\author{Angel Stanoev} 
\affiliation{Macedonian Academy for Sciences and Arts, Skopje, Macedonia \\ 
Email: astanoev@cs.manu.edu.mk}

\author{Daniel Smilkov} 
\affiliation{Macedonian Academy for Sciences and Arts, Skopje, Macedonia \\ 
Email: dsmilkov@cs.manu.edu.mk}

\author{Ljupco Kocarev}
\affiliation{Macedonian Academy for Sciences and Arts, Skopje, Macedonia \\
 BioCircuits Institute, University of California, San Diego \\
 9500 Gilman Drive, La Jolla, CA 92093-0402 \\ Email: lkocarev@ucsd.edu}

\title{Identifying communities by influence dynamics in social networks}

\begin{abstract}

Communities are not static; they evolve, split and merge, appear and disappear, i.e., they are product of dynamical processes that govern the evolution of the network. A good algorithm for community detection should not only quantify the topology of the network, but incorporate the dynamical processes that take place on the network. We present a novel algorithm for community detection that combines network structure with processes that support creation and/or evolution of communities. The algorithm does not embrace the universal approach but instead tries to focus on social networks and model dynamic social interactions that occur on those networks. It identifies leaders, and communities that form around those leaders. It naturally supports overlapping communities by associating each node with a membership vector that describes node's involvement in each community. This way, in addition to the overlapping communities, we can identify nodes that are good followers to their leader, and also nodes with no clear community involvement, that serve as a proxy between several communities and are equally as important. We run the algorithm for several real social networks which we believe represent a good fraction of the wide body of social networks and discuss the results including other possible applications.


\end{abstract}


\maketitle

\section{Introduction} \label{sec:intro}

Biological, technological and social complex systems are networked: their structure can be represented as networks of interacting components. This makes networks a very powerful tool for understanding the structure, dynamics and evolution of complex systems \cite{newman-book}. Very often these networks exhibit modular and hierarchical structure that supports their evolution into a highly complex systems \cite{hierarchy1,hierarchy2,hierarchy3}. The automatic detection of this modular structure -- also known as community detection -- can help identify closely related class of nodes and give a principled way of understanding the organization of complex systems \cite{fortunato2010}.

However, current research for community detection focuses on finding algorithms that can identify communities in all contexts \cite{clique, modularity2, infomap}. This universal approach has many drawbacks, the most important being that these algorithms fail to explain the produced partition. In order for algorithms to be usable in practical contexts, we need to incorporate context-based knowledge about how communities are built and how they evolve. For example, in social networks the communities are usually built around some important individuals or group of individuals called leaders. In communication networks, modules are built around highly connected hubs and in paper citation networks, communities correspond to the different research areas and important papers in those areas.

Also, communities are not static, they evolve, split and merge, appear and disappear, i.e., they are product of dynamical processes that govern the evolution of the network \cite{jure}. Therefore, a good algorithm should not only quantify the topology of the network, but incorporate the dynamical processes that take place on the network as well. Since there are many dynamical processes: spreading diseases, packet routing, viral marketing, random walks, consensus dynamics etc, it is difficult to produce an algorithm that will perform well on every complex network. Communities in networks often overlap such that nodes simultaneously belong to several groups \cite{clique,overlap}. Surprisingly however, this property has been continuously disregarded until recently \cite{link-community,landscapes}, where few algorithms have been introduced, but their number is still substantially smaller than the number of non-overlapping algorithms.

In this paper we present a novel algorithm for community detection with focus on social networks. The algorithm does not embrace the universal approach but instead tries to focus on social networks and model the dynamic social interactions that occur on those networks. This helps to identify leaders and communities that form around those leaders. It naturally supports overlapping communities by associating each node with a membership vector that describes node's involvement in each community.

The outline of the paper is as follows. Section \ref{sec:social-networks} introduces the problem of community detection in social networks by using the famous Zachary social network as a case study to explain the problems with the current approaches and the motivation behind our algorithm. In section \ref{sec:algorithm} we present our algorithm and explain its steps. In section \ref{sec:applications} we run the algorithm for several real social networks which we believe represent a good fraction of the wide body of social networks. We also discuss other possible applications. Section \ref{sec:conclusion} concludes this paper.

\section{Motivation} \label{sec:social-networks}

The most important part of a social network are its ties, or connections, that denote some kind of social relationship. We believe that a simple quantification of these connections has many drawbacks. A good method for community detection must rather focus on the social relationship than on the bare connection, i.e., it must focus on processes that support these connections and the creation of communities. In this section we use the Zachary social network \cite{zachary} as a case study to discuss some of the drawbacks of current methods. We also explain our motivation behind the proposed algorithm.

The most popular methods for community detection are based on modularity quality function. Since its introduction \cite{modularity-intro}, there have been many community detection algorithms that use the modularity function as a basis \cite{modularity2,modularity3,modularity4}. These algorithms usually optimize this function in order to achieve a greater modularity value as a result, and consequently a better community detection. But recently the focus on the modularity function seems to be lost, mainly because of the shortcomings of the function discovered. Among others, the two most important are the resolution limit of the modularity function \cite{modularity-dback,modularity-dback2,modularity-dback3,modularity-dback4} and the structural diversity of high-modularity partitions \cite{modularity-dback5}. Basically, the optimal partition may not coincide with the most intuitive partition.

\begin{figure}[htbp]
  \begin{center}
   \includegraphics[scale=0.4]{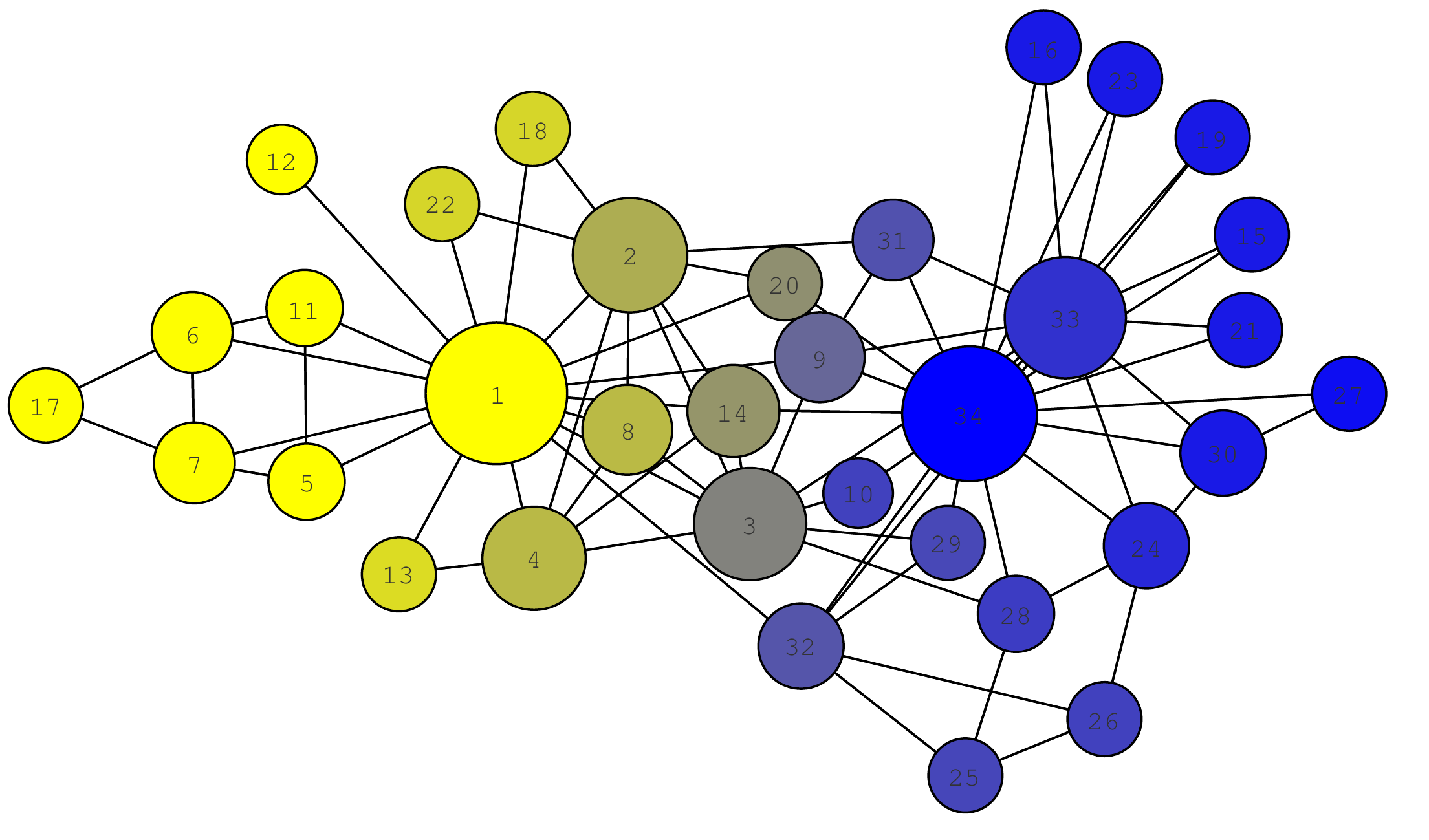}
  \end{center}
  \caption{(Color online) The Zachary karate network \cite{zachary}. Leaders with id 1 and 34 form communities and spread their influence through the network. The partition found by our algorithm not only matches the original partition, but also identifies the exact leaders.}
  \label{fig:zachary}
\end{figure}

We found another shortcoming of the modularity function on border case nodes (see Appendix \ref{app:misjudgement}). Let's look at the Zachary social network with focus on node with id 10 (Fig. \ref{fig:zachary}). We will ignore the coloring of the nodes for now. Since this is a social network, we should consider the social dynamics that are taking place, namely the influence spreading over the network. Node 10 has two neighbors, node 34, which is denoted by the author as a leader in the first community, and node 3, which is neighbor of node 1, the leader in the second community. Clearly, node 34 has more influence in the network then node 3, so for example, if elections are being held in the karate club, node 10 will most probably vote for node 34, than for node 1. Consequently, node 10 should belong to the first community where node 34 is the leader. This emphasizes the idea that the assignment method should take into account the dynamics, not only the topology. On the other hand, modularity function produces greater value when node 10 is in the second community, and that decision is made only because the second community has smaller number of links (see Appendix \ref{app:misjudgement}). All modularity-based algorithms will fail to produce the right partition of the network, since they are driven only by the network topology. There is also an implicit hierarchy in this network. There are 2 leaders and communities are build around those leaders. The removal of those leaders will result in splitting these communities since leaders are keeping these communities together. Identifying the leaders will implicitly result in identifying the communities. Furthermore, to avoid the well-known resolution limit, decision-making process should be made on node level, and not globally on the whole network. Today we have very large, but sparse networks. That is why we believe that the decision in which community a node should belong, should be based on the node's neighborhood solely. We have networks that are growing rapidly fast, but the node's horizon is not growing beyond its neighborhood.

\begin{figure}[htbp]
  \begin{center}
   \includegraphics[scale=0.7]{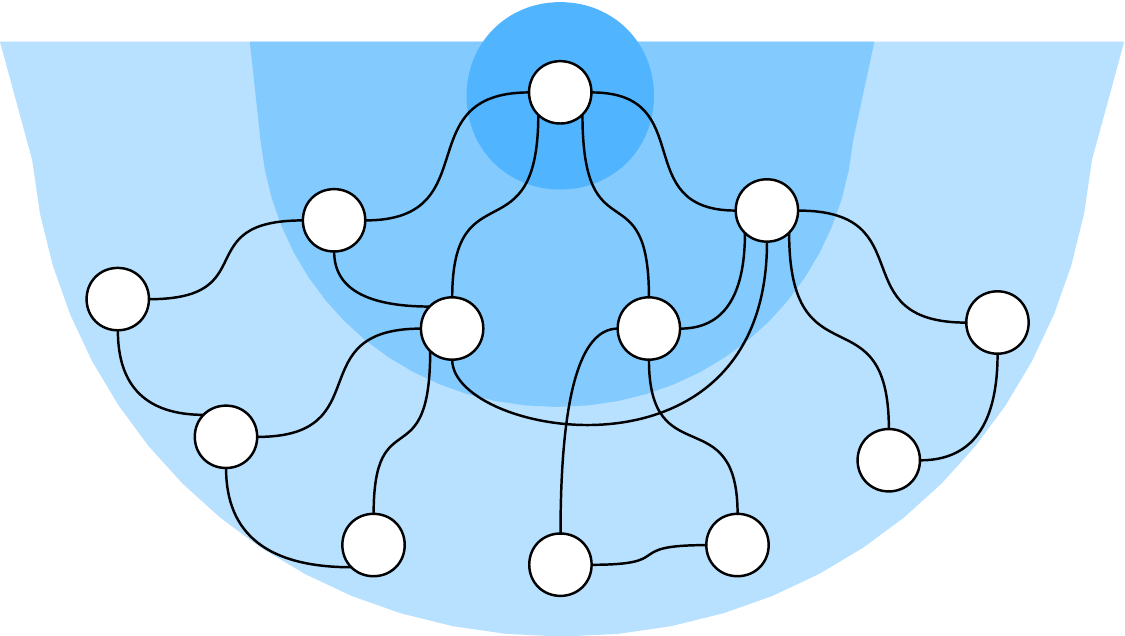}
  \end{center}
  \caption{(Color online) Social hierarchy within a community. The more influential nodes are located on a higher level in the hierarchy. The leader is located on the highest level. Semicircles depict different levels in the hierarchy with the darkest color denoting the highest level.}
  \label{fig:hier}
\end{figure}

When talking about influence, it is natural to talk about hierarchy as well. In one such hierarchy there are nodes that are more important and influential than some other nodes, hence located on a higher level in the hierarchy. It naturally follows that the leader is located on the highest level within that hierarchy (see Fig.\ref{fig:hier}). Since the hierarchies are consequence of the spreading of influence, and so are the communities, we believe that the identification of these hierarchies in a network will result in a natural community detection. The area in which a leader has most influence should define its community. So, community detection is performed by finding all natural leaders and all nodes on which they influence. Partitions obtained this way can be naturally explained. Also, another intuitive property that a community should possess is satisfied this way, which is the property that shortest paths between nodes from a same community should consist only of nodes from that community. With well defined hierarchies, the shortest paths will be approximately the paths that run through the hierarchical trees.

\section{The algorithm} \label{sec:algorithm}

To sum up, the hierarchical point of view and its significance, the natural community detection from it and the need of decentralized approach, are the basis of our algorithm. The first step is to define the amount of influence a node has on another node. Real networks have significantly high clustering coefficient, meaning the nodes tend to form triangles with other nodes. Here, we make use of the idea that the link density is greater within a community than between the communities. That means that more triangles are formed in the communities, than outside the communities.
In \cite{genetic}, interesting characteristics about a node's social embeddings are discovered. The in-degree can be explained by the person's genes in 46 percent of the cases. But a more non-obvious characteristic is that 47 percent of the variation, whether a person's friends know one another, is attributable to the person's genes. 
Some people like to introduce their friends to each other and form communities around them, and others simply do not do that. And that is what separates the leader in a group from a regular person in that group. The leader tends to connect its neighbors with one another in order to build a stronger community around it, whereas a more margin person is more of a subject of being connected to someone, be a member of something, rather than connect someone, or create something, or influence someone.
Therefore, if a node can find the ``direction'' where the most of its triangles are placed, then its community is also in that ``direction''. The denser the triangles are, the closer the node is to the core of the community. We also believe that triangles between two neighbors serve as better proxy for influence than just the direct connection between the nodes. So it is natural that the more triangles a neighbor shares with a node, the more influence it has on that node. This shows the connection between the influence dynamics and the topology of the network measured with triangles.

In the following, we focus on simple directed weighted network $G$ with no multiple links and self-loops, described by its $N\times N$ adjacency matrix $A$, where $N$ is the number of nodes. By definition, $A_{ij}$ is the topological weight of the link going from $j$ to $i$ and $A_{ij}\neq A_{ji}$ in general. Also, since dealing with directed networks, we interpret $A_{ij}$ as proportional to the influence (trust) node $i$ (node $j$) has on node $j$ (node $i$). If the network is undirected $A_{ij}=A_{ji}$. $s_i=\sum_j{A_{ij}}$ is the strength of node $i$ and when the network is unweighted, $s_i$ is simply the in-degree of node $i$.

\subsection{Influence matrix}

\begin{figure}[htbp]
  \begin{center}
  \includegraphics[scale=0.5]{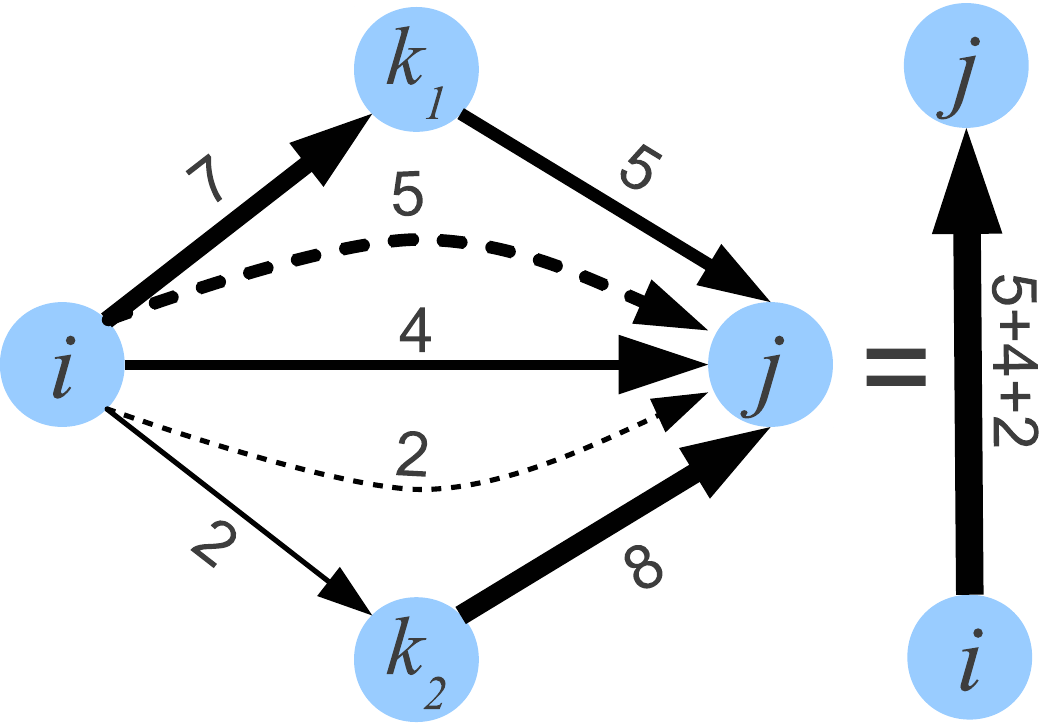}
  \end{center}
  \caption{(Color online) Embedding triangle information into the link weights. Solid lines depict ``direct'' links and dashed lines depict ``transitive'' links. The ``transitive'' link weights are obtained as the minimum weight of the links of which are deducted. Line labels denote link weights with width proportional to their weight.}
  \label{fig:transform}
\end{figure}

To incorporate the information we have from triangles, we introduce network $G'$, a weighted network where triangles are embedded into the link weights, thus obtaining the influence matrix $A'$. To do this, let $C_{ji}^k=\min \{A_{ki},A_{jk}\}$ be the ``transitive'' link weight from node $i$ to node $j$ through node $k$. We define this only for neighboring nodes, thus $C_{ji}^k=0$ if $A_{ji}=0$. We choose the minimum of the two link weights motivated by the expression ``a chain is only as strong as its weakest link''. Together with the ``direct" link weight $A_{ji}$, we obtain the new link weight $A'_{ji}=A_{ji}+\sum_k C_{ji}^k=A_{ji}+\Delta_{ji}$. This procedure is illustrated in Fig. \ref{fig:transform}. If the network is undirected and unweighted, $A'_{ji}$ is simply the number of triangles between neighbors $i$ and $j$ plus $1$. Also, for later use, we will define now $\Delta_j=\sum_k\Delta_{jk}$ which for undirected and unweighted network is simply twice the number of triangles containing node $j$ . However, if link weights represent the actual influences we are trying to extract (including the transitive weights), this step should be omitted, i.e., one can take $A'_{ji}$=$A_{ji}$.


The next step is calculating the \textit{overall influence} $x^*_i$ of every node $i$ in the network. The overall influence $x^*_i$ represents how important is the opinion of node $i$ in the network, i.e., how much its opinion spreads through the network. As a process for modeling the influence spreading in the network, we consider the unbiased random walk where, at each step, a walker at node $j$ follows one of the outgoing links proportionally to the link's weight $A'_{ij}$. Writing $\mathbf{x}(t)=[x_1(t)\:x_2(t) \ldots x_N(t)]$, where $x_i(t)$ is the  overall influence of the node $i$ at time $t$, the expected density of walkers evolves according to the rate equation
\begin{equation}
\label{eq:random-walk}
\mathbf{x}(t+1)=T\mathbf{x}(t)
\end{equation}
where $T$ is the transition matrix whose entry $T_{ij}$ represents the probability to jump from $j$ to $i$,
\begin{equation}
\label{eq:relat-influece} 
T_{ij}=\frac{A'_{ij}}{\sum_kA'_{kj}}.
\end{equation}
$T_{ij}$ denotes the \textit{relative influence} node $i$ has on node $j$. We start with initial vector $\mathbf{x}(0)=[\frac{1}{N}\:\frac{1}{N}\ldots\frac{1}{N}]$. 
The overall influences $\{x^*_i\}$ is a steady-state solution of  (\ref{eq:random-walk}) and can be obtained for directed networks only numerically by iterating (\ref{eq:random-walk}), that is, when time $t$ goes to infinity.  In a special case when the network is undirected and non-bipartite, there is a known analytical solution for $x^*_i$, i.e.,  
$$x^*_i=\sum_jA'_{ij}/\sum_{jk}A'_{kj}\propto s_i+\Delta_i. $$
Note that node's potential of becoming a leader depends on the in-degree and number of triangles, as discussed earlier in this section.

\subsection{Leaders identification}

Since we now know the relative influences between the nodes $T_{ij}$, and the overall influences of nodes $x^*_i$, we can find the leaders in the network. 
%
%
A leader should have big overall influence, since the overall influence represents how close a node is to the core of its community, and the actual potential of becoming a leader. Also, a leader should have more influence on its neighbors than they have on it. 
Therefore we define leaders as those nodes for which the product (overall influence) $\times$ (relative influence) is large. More precisely,  
we denote with $\Gamma_i=\{j | T_{ji}=\max_k{T_{ki}}\}$ the set of neighbors with the largest relative influence on node $i$. 
Node $i$ is a \textit{leader} if:
\begin{equation}
T_{ij}\cdot x_i^*>T_{ji}\cdot x_j^*
\label{influence} 
\end{equation}
for all $j \in \Gamma_i$. The product $T_{ij}\cdot x_i^*$ of two numbers $T_{ij}$ and $x^*_i$ combines the relative influence of node $i$ towards node $j$ with the overall influence of node $i$.

Note that in the rare cases where two or more leaders are also most influential neighbors between each other, (that is, when $T_{ij}\cdot x_i^* = T_{ji}\cdot x_j^*$ ), than they are grouping together and are becoming leaders of one group. For example, in a full mesh network, all of the nodes are leaders of one community, whereas for a ring network, each node is a leader to its own community. Actually, this suggests that in the cases where there is a lack of hierarchical structure, no particular leader in a community, the community will be split on subgroups and the partition will depend on its link density.

\subsection{Computing the membership vectors}

Suppose we have $L$ leaders in the network, hence $L$ communities and let $l=\{l_1,l_2, \ldots, l_L\}$ be the set of all the leaders. We calculate the membership vector $\mathbf{y}_i=[y_i^1\:y_i^2 \ldots\:y_i^L]^T$, a probability vector of length $L$, that describes node $i$'s involvement in each community. Since  $\mathbf{y}_i$ is a probability vector, its components sum to 1, i.e. $\sum_{k=1}^L y_i^k = 1$. For every leader $l_i$, the initial membership vector $\mathbf{y}_{l_i}(0)$ has all the components equal to zero, except for the $i$-th component $y_{l_i}^i=1$. For every node $j$ that is not a leader, all the components of $\mathbf{y}_j(0)$ are initialized to $\frac{1}{L}$ to denote equal participation to each community. For computing the membership vectors, we consider consensus dynamics, i.e.
$$\mathbf{y}_i(t+1)=\frac{1}{\sum_jA_{ji}}\sum_j A_{ji} \mathbf{y}_j(t)=\sum_j S_{ji}\mathbf{y}_j(t).$$
At each time step, the membership vector of each node is updated by computing a weighted average of the membership vectors of its neighbors. We do not use matrix $A'$ since the influence embedded in $A'$ will naturally occur in this process and its inclusion can introduce bias. However, if $S$ is irreducible, which is often true for undirected graphs, this system will converge to a consensus state, where all the nodes reach an agreement, thus having the same membership vector. To avoid this, we keep the leader's membership vector immutable, i.e.,
$$\mathbf{y}_{l_i}(t+1)=\mathbf{y}_{l_i}(t)= \ldots =\mathbf{y}_{l_i}(0).$$
This way, we modify matrix $A$, by connecting each leader only by itself. After this modification, matrix $S$ remains a Markov matrix with every column summing to 1, which guarantees convergence since the largest eigenvalue of $S$ is 1 and its multiplicity is $L$.

\section{Applications to real networks} \label{sec:applications}

\subsection{Properties of the algorithm} 

In this subsection we discuss four properties of the algorithm: detecting overlapping nodes, detecting leaders, detecting hierarchical organization, and decentralization.

\subsubsection{Detecting overlapping nodes}

An important property of our algorithm is the computation of a membership vector for each node. Instead of having one number denoting its membership in a single community, we have a percentage for each community. As a result, we can easily identify nodes that naturally belong to more than one community, known as overlapping nodes. Additionally, we can find nodes that are good followers of their leader, but also nodes that have no distinguished leader and serve as a proxy between several communities.

\subsubsection{Detecting leaders}

As our algorithm is best suited for real networks where the process of influence spreading takes place, it is only natural that it can be used for influence related problems. One such, is the actual identification of the leader in a community. By detecting the leader in a community we gain very useful information, as the leader, by the definition of the algorithm, is the most influential node in its community. By removing the leader it can be expected for the community to suffer serious consequences, like splitting up on several smaller communities or a complete degradation. The leader's hierarchy, or the leader's community, is the area where the leader's opinion is the most influential opinion. This can be used for an efficient viral marketing campaign, for example. One interesting feature of the algorithm is that although it automatically detects the best leaders, one can specify a priori some nodes as leaders and build community structures around them.

\subsubsection{Detecting hierarchical organization}

Another characteristic feature of the algorithm is the possibility of deriving the hierarchical organizations of the communities. A node's parent can be easily detected by the influence matrix and the overall influences. It can be the most influential neighbor of its community, or it can be the most strictly oriented neighbor towards the same community, actually on a higher hierarchical level.
This can be used in communication networks, where a node can use a hyperbolic greedy algorithm to forward packets to other nodes in the community \cite{rkleinberg}, which is important since the communication is more frequent within a community. As for the other nodes, the greedy algorithm can be used to forward the packets to the leader, supposing that the leader knows how to forward those packets to its respective leader. Also, a node's siblings can be detected, as they all share the same parent. This may be used in prediction of missing links scenario, for example.

\subsubsection{Decentralization}

The idea behind decentralization is that a node should be able to decide in which community it belongs only by considering its neighborhood, without taking into account any global characteristics of the system. An important property of our algorithm is that it can be applied on decentralized scenarios. In the first step of the algorithm, we only need the connectivity in the neighborhood of each node to determine the matrix $A'$. In the second step, the influences are computed with random walk iterations which can run in distributed fashion using message passing. Leaders identification involves a direct message exchange between everu node $i$ and its potential neighboring leaders ($\Gamma_i$). In the last step, the leaders spread their influences in the network through their neighbors with message passing and within several iterations, the system stabilizes to the desired state. All of the described steps can be carried out in a decentralized fashion with message passing. Also, our algorithm can incorporate network dynamics as well. If a new node is added to the network, it finds its parent and calculates its membership vector. If a node is removed from the network, or a link is added/removed from the network, the affected nodes can detect their parents and recalculate their membership vectors. 


Even though designed with social networks in mind, we believe our algorithm can be used in various contexts. Very often, in wireless sensor networks with low energy requirements and limited sensor memory, we need to aggregate the sensor data of the nodes. Since the nodes are being deployed in an Eucledian space, one should expect non-negligible number of triangles. Also, we can expect the detected communities to depend on the geographic distribution of the nodes (smaller communities to be detected with approximately equal sizes if the geographical distribution of the nodes is uniform). Furthermore, the sensor data aggregation is geographically-based. As a result, the aggregation on a community level will be a good aggregation. The hierarchical organization within a community can be very useful for the aggregation process. Clearly, the leader is best suited to be an aggregator, so the nodes should transfer their sensor data to the leader. Even more, one can assign arbitrary nodes as aggregators, such as nodes that have more resources. Since the sensor nodes have very limited resources (processing, memory, energy etc.), a simple memory-free hyperbolic greedy algorithm, based on the derived hierarchy, can be of great significance \cite{rkleinberg}.


If executed in centralized fashion, our algorithm has low computational complexity varying from $O(N)$ to $O(N^2)$, depending on the power-law exponent of the degree distribution and the number of detected communities (leaders) (see Appendix \ref{app:complexity}).

\subsection{Real-world networks}

In order to verify the validity of our algorithm, we run the algorithm for several real social networks which we believe represent a good fraction of the wide body of social networks. The networks are small, easy to visualize and have been explored by many researchers studying social behavior. Thus, we can visually and verbally measure the performance of our algorithm. Also, our \textit{algorithm is fast enough to work with large networks having millions of links}, but we did not find a social network rich enough with meta-data to objectively measure our algorithm. Furthermore, we avoid the LFR benchmark since its connection to real social networks is questionable and if we ignore that, still we will be unable to validate if our algorithm found the real leaders. When visualizing the results, each leader is assigned different color and each node is assigned a color which is a weighted average of the colors of the leaders in the network based on the membership vector. When visualizing, the layout is done by the Fruchterman-Reingold algorithm \cite{FR}, with node sizes proportional to their overall influences. When we compare partitions, we marginalize our result by assigning each node to the community with the highest component in the corresponding membership vector. This can also be seen visually as each node has a dominant color. 

\subsubsection{The Zachary karate network}
One of the most popular networks for validating community structures is the Zachary karate network \cite{zachary}. It is a friendship network consisted of social interactions among members of a karate club, so it is driven by spreading of influence. The author denoted two nodes as leaders in the network, the president of the club and the instructor (node 34 and 1 respectively), and two respective communities. The communities have been created after a drift between the two leaders (the president and the instructor of the club). As it is shown on Fig. \ref{fig:zachary}, the partition found by our algorithm not only matches the original partition, but also identifies the exact leaders.

\subsubsection{Bottle-nose dolphins network}
\begin{figure}[htbp]
  \begin{center}
  \includegraphics[scale=0.3]{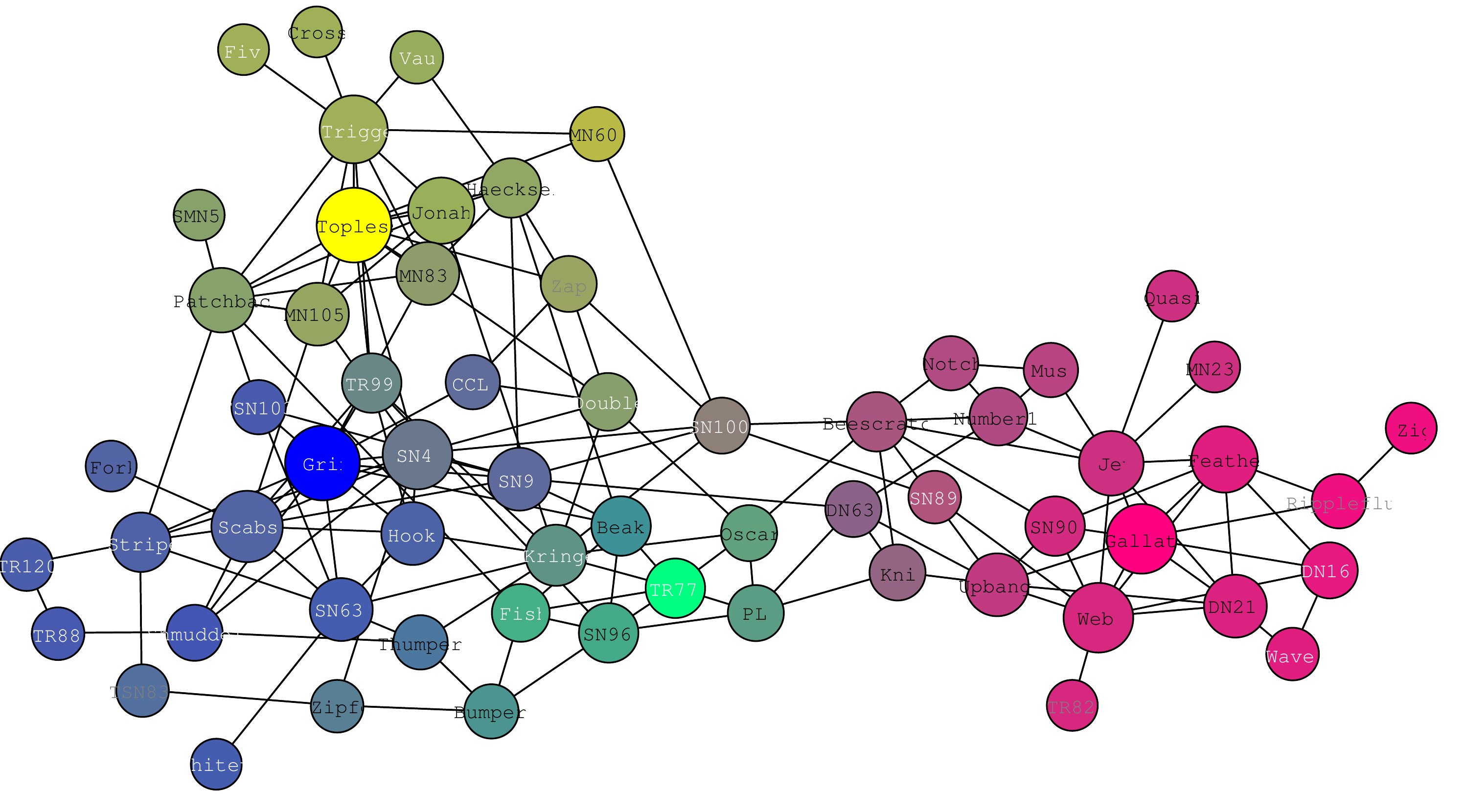}
  \end{center}
  \caption{(Color online) The Bottle-nose dolphins network. Four leaders are detected: Topless, Grin, TR77 and Gallatin. Gallatin's community and the combination of the other three communities gives the main strong division, noted by the authors. Almost all of the Grin-oriented dolphins are female, and most of the other dolphins are male. The female dolphins are labeled with white, the male with black and those with unknown gender are labeled with gray color. There is clear statistically significant assortative mixing by sex among the dolphin population \cite{dolphins3}, and also access to oestrus females tends to be the main driver of male sociality \cite{dolphins4}, which in a way explains our partition.}
  \label{fig:dolphins}
\end{figure}

Another popular real-world network in the community detection field is the bottle-nose dolphins of Doubtful Sound network \cite{dolphins1,dolphins2}. The network consists of 62 dolphins observed in a seven years period, with links corresponding to significant frequent association. The network was split into two communities \cite{dolphins3} for the period when a dolphin (node SN100) located between the groups temporarily disappeared. Further, there was also detected clear statistically significant assortative mixing by sex among the dolphin population. So, for this network we only have a predefined strong division, with further divisions probably dependent on the gender. Our algorithm detects four communities (see Fig. \ref{fig:dolphins}) where if Topless's, Grin's and TR77's communities are combined, we have the original strong division into two groups. The produced partition is very similar to the ones produced by \cite{dolphins3} and \cite{modularity-intro}. Most of the Topless-, TR77- and Gallatin-oriented dolphins are males (black labels), and almost all of the Grin-oriented dolphins are females (white labels). There are four nodes with unknown gender (gray labels). Access to oestrus females (females in rutting season) tends to be the main driver of male sociality \cite{dolphins4}. The evolution of the complex relationships between male groups was driven by sexual competition probably to out-compete other males for female choice. Topless's and TR77's communities seem to be driven by those rules as well, as the author noted that most of the males from those groups spent significantly more time with oestrus females than with the other female groups. Indeed it can be seen that the male dolphins from Topless's community have significant association with the female dolphin Trigger and the females from Grin's community. The core dolphins from Gallatin's community did not spent significantly more time with the oestrus females. In a way, this confirms our partition as a good one. Some of the detected leaders are identified as central individuals by the author \cite{dolphins2}.

\subsubsection{Sawmill network}
\begin{figure}[htbp]
  \begin{center}
  \includegraphics[scale=0.4]{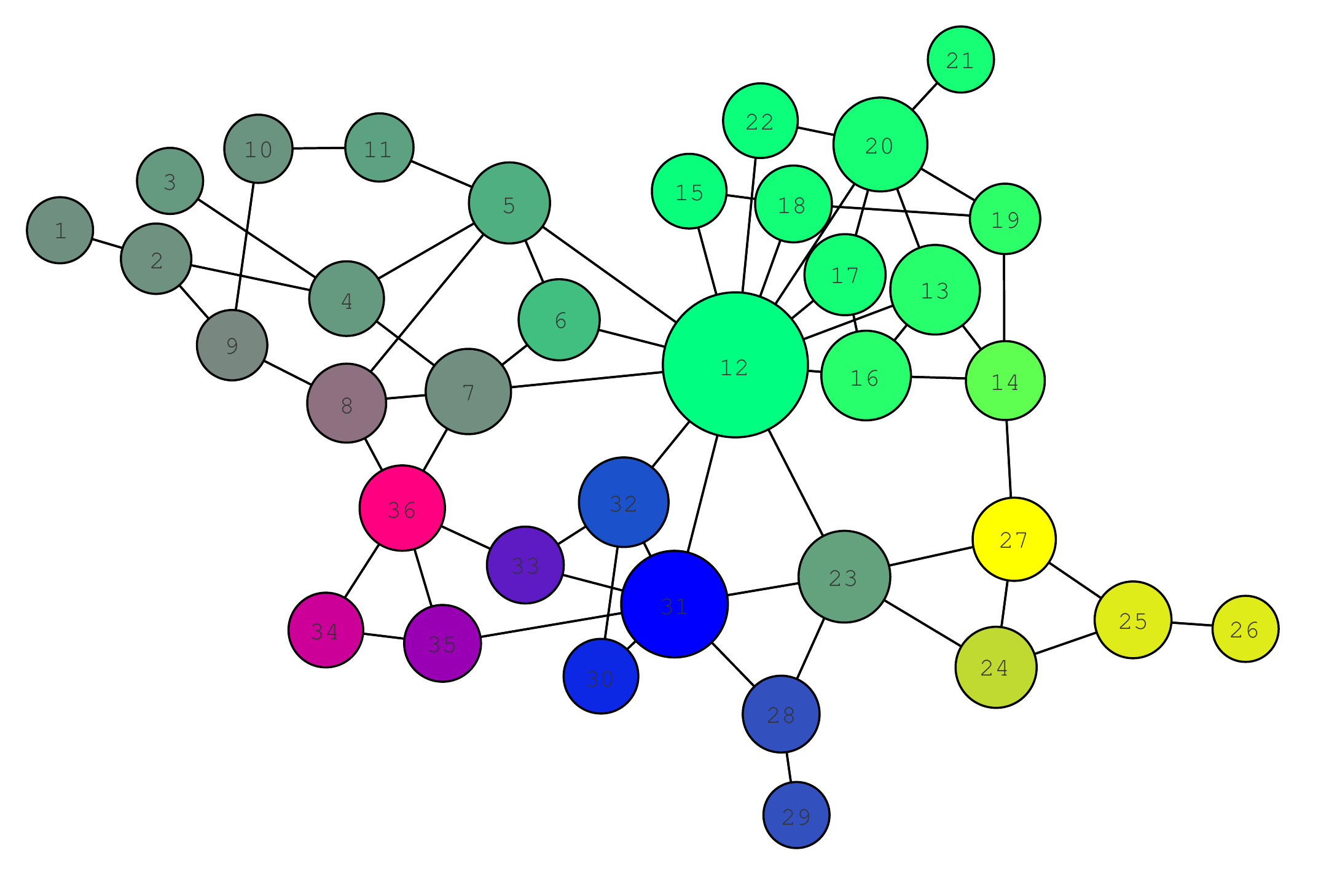}
  \end{center}
  \caption{(Color online) The Sawmill communication network. Our algorithm successfully identifies the sectors within the sawmill and the divisions corresponding to the native languages, with only difference being the merging of the two Spanish sectors. That comes as a result of the lack of hierarchical and community structure in one of the sectors and the big overall influence of the employee Juan (node 12), also noted by the author \cite{sawmill}.}
  \label{fig:sawmill}
\end{figure}

Fig. \ref{fig:sawmill} shows the sawmill communication network, which is a communication network between the employees within a sawmill \cite{sawmill}. The network consists of employees speaking English and Spanish language. Also, there are four sectors, the planer crew, the mill crew, the mill management and the yard. There are two non-sector members - the kiln operator and the forester. The large sectors - the planer crew and the mill crew - are further divided into two subgroups corresponding to the native language. Our algorithm detects four communities with four leaders: nodes 12, 36, 31 and 27. Two of the communities correspond to the English planer and mill crew, node 36's and node 27's, respectively. The Hispanic planer and mill crew (Spanish native) are joint together in node 12's community. This comes as a result of the lack of hierarchical and community structure in the planer crew (up-left), meaning none of the nodes act as a leader. As a consequence, the nodes of this group are mostly oriented towards the employee Juan (node 12), which is due to the big overall influence that the employee Juan has. A significant information flow is conducted through that employee as noted by the author as well. Also, the nodes from the Hispanic planer crew are strongly influenced by node 36, which is a leader of the English planer crew, so their colors are mixture of node 12's and node 36's colors. The final community is the mill management, merged with the small Yard sector (only two employees), the kiln operator and the forester - node 31(mill manager)'s community.

\subsubsection{Sawmill strike network}
\begin{figure}[htbp]
  \begin{center}
  \includegraphics[scale=0.5]{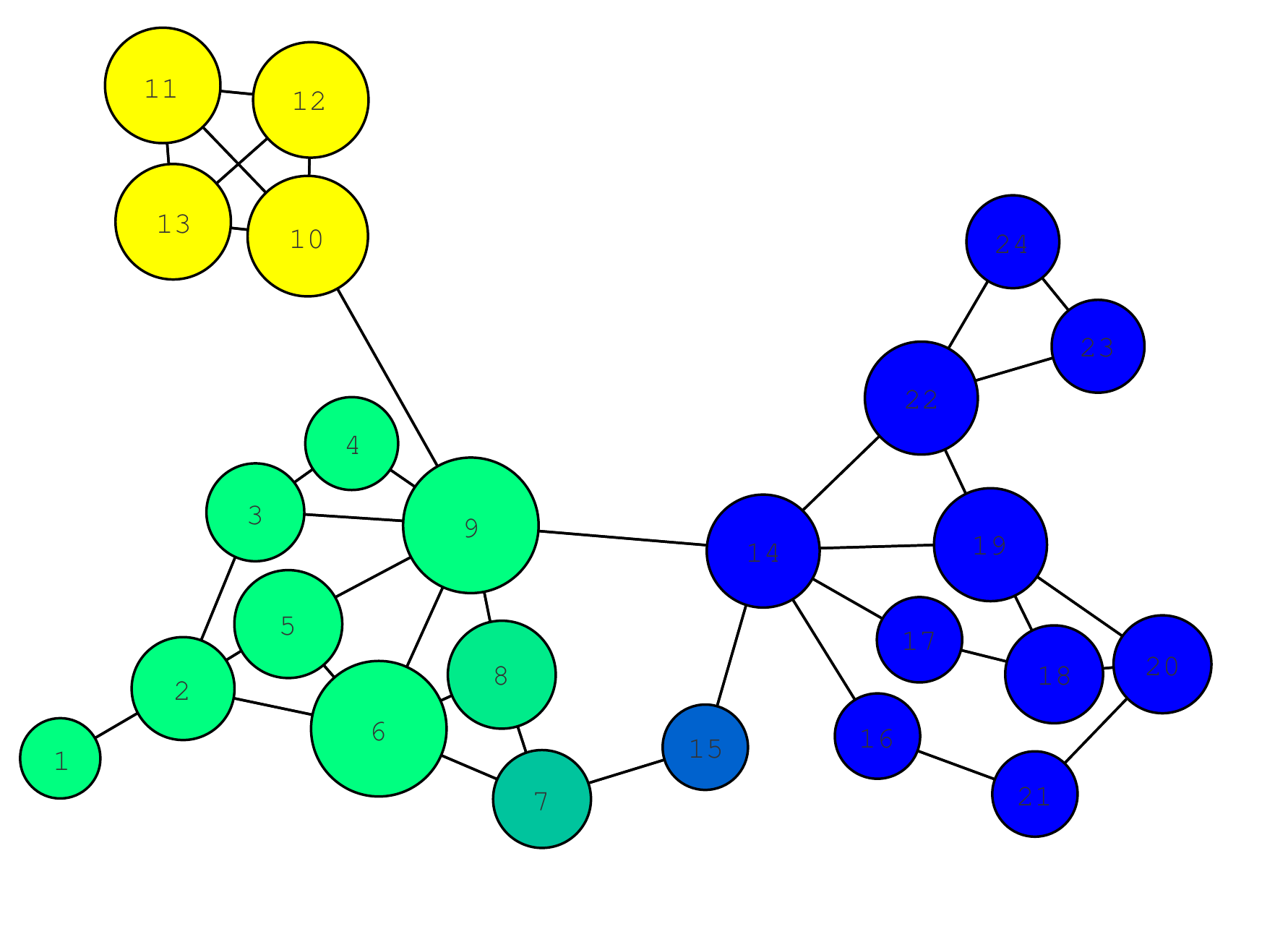}
  \end{center}
  \caption{(Color online) The Sawmill strike network. Communication network of the employees within a sawmill during a period of a strike \cite{strike}. The network has three communities, also correctly identified by our algorithm: young English group (bottom-left), young Spanish group (top-left) and old English group (right). The leaders detected by the algorithm are also noted by the author as most central nodes in their groups.}
  \label{fig:strike}
\end{figure}

Fig. \ref{fig:strike} shows the communication network between the employees within a sawmill during a period of a strike \cite{strike}. The strike occurred as a result of the new rules, installed by the new management, that changed the workers' compensation package. Company management (not shown in the figure) perceived that the two union negotiators were not fully communicating their terms with all of the union members. They felt that the new wage package was not being properly explained to all employees by the union negotiators. The research reveals the network structure. There exist two groups according to age division (see Fig. \ref{fig:strike}) - a group of older employees (over 38 years old - right side) and a group of younger employees (under 30 years old - left side). Further, in the group of younger employees there is a division due to the native language - English (bottom) and Spanish (top). The author denoted the nodes with id 9 and with id 14 as the most central nodes in the young and old group, respectively. The same are identified as leaders by our algorithm as well. The node with id 10 is also identified as a leader, and is noted by the author as the most proficient English speaker from the densely-connected Hispanic group, and the only one that communicates outside that group. The research helped in the resolving of the negotiations stalemate between the new management and the negotiators (nodes with id 22 and 24). Since the main problem was perceived to be the lack of communication between the negotiators and the rest of the employees, particularly the young ones, a cooperation with the nodes 9 and 14 was proposed, so the communication would be improved. That was the actual case, as the more than 3 weeks old strike was ended within 48 hours, and the production was restarted shortly thereafter.

\section{Conclusion} \label{sec:conclusion}

After the discovering of the main drawbacks of modularity function \cite{modularity-dback,modularity-dback2,modularity-dback3,modularity-dback4,modularity-dback5}, its focus among researchers has been slightly decreased, and we are expecting a new wave of different approaches and algorithms in the next years. Our algorithm is following this flow by not embracing the universal approach, but rather focusing on social networks with the dynamic social interactions that occur on those networks. The membership vectors found by our algorithm are much more descriptive than a partition; we obtain partitions by marginalizing the membership vectors. Besides community detection, identifying leaders can be very important when modeling dynamics between a group of opposing members in a network, such as elections and marketing campaigns.

\appendix
\section{Misjudgment of the modularity function} \label{app:misjudgement}

Suppose we have a graph $G$ with $n$ nodes, with two communities $C_1$ and $C_2$, and with total of $m$ links between the nodes. We observe a single node $x$ in the network. It has $d_1$ links to nodes from the $C_1$-community and $d_2$ links to nodes from the $C_2$-community. We want to know how the modularity function makes the decision on whether it places the node $x$ in the $C_1$-community or in the $C_2$-community. That is actually how the network topology, i.e., the communities' sizes and number of links and nodes, influence the value of the modularity function for a given partition. Let $Q^1$ be the modularity value if the node $x$ is placed in the $C_1$-community and $Q^2$ be the value if it is placed in the $C_2$-community. Let $Q^1_x$ be the contribution that the node $x$ gives to the partition with the joining of the $C_1$-community, and $Q^2_x$ the contribution of joining the $C_2$-community.
The modularity function is given by
$$Q=\frac{1}{2m}\sum_{ij}\left[A_{ij}-\frac{k_ik_j}{2m}\right]\delta(c_i,c_j)$$
where $c_i$ is the community to which node $i$ belongs, $\delta(c_i,c_j)$ is the Kronecker delta symbol, $k_i$ is the degree of node $i$. We take $K_i=\sum_{j \in C_i, j \not= x}k_j$. So
\begin{eqnarray*}
Q_x^1 &=& \frac{1}{2m}\sum_j\left[A_{xj}-\frac{k_xk_j}{2m}\right]\delta(C_1,c_j) \\
&=& \frac{1}{2m}\left[d_1-\frac{k_x}{2m}\sum_{j \in C_1, j \not= x}k_j\right] \\
&=& \frac{1}{2m}\left[d_1-\frac{d_1+d_2}{2m}K_1\right].
\end{eqnarray*}
In a similar way, 
$$Q_x^2=\frac{1}{2m}\left[d_2-\frac{d_1+d_2}{2m}K_2\right].$$
Since 
\begin{eqnarray*}
Q^1-2Q^1_x & = & \frac{1}{2m}\sum_{ij,i\not=x,j\not=x} \left[A_{ij}-\frac{k_ik_j}{2m}\right] \delta(c_i,c_j) \\ &=& Q^2-2Q^2_x
\end{eqnarray*}
we have
$$Q^1<Q^2 \Leftrightarrow Q^1_x<Q^2_x.$$
We want to know when the modularity function will choose $C_2$ over $C_1$, so we will explore the value of $Q^1_x<Q^2_x$
$$Q^1_x-Q^2_x=\frac{1}{2m}\left[d_1-d_2-\frac{d_1+d_2}{2m}(K_1-K_2)\right].$$
If we assume $d_1=td_2$, where $t \geq 1$ is an integer, we have
$$Q^1_x-Q^2_x=\frac{1}{2m}\left[(t-1)d_2-\frac{(t+1)d_2}{2m}(K_1-K_2)\right]$$
In order for this expression to be smaller than 0, we must have
$$(t-1)d_2<\frac{(t+1)d_2}{2m}(K_1-K_2),$$
that is 
$$(t-1)2m<(t+1)K_1-(t+1)K_2.$$
Since $K_1+K_2+d_1+d_2=2m$, we obtain 
$$(t-1)(K_1+K_2+(t+1)d_2)<(t+1)K_1-(t+1)K_2,$$
that is 
$$2tK_2+(t^2-1)d_2<2K_1.$$
For example, let us consider the simplest case: the node $x$ has one link to a node from the $C_1$-community and one link to a node from the $C_2$-community. So, $t=1$ and $d_2=1$. We have that $K_2<K_1$
is the only condition for the modularity function to choose the $C_2$-community. This is the exact case for the Zachary karate club network and the node with id 10. The modularity function produces greater value when the node 10 is placed in the community of the node 1, only because that community is smaller (links-wise) and it does not give any significance to the fact that one of the neighbors of the node 10 is the leader of the other community, the node 34.
As another example, let the node $x$ have t links to nodes from the $C_1$-community and 1 link to a node from the $C_2$-community. So, $d_2=1$, and we have that $2tK_2+t^2-1<2K_1$ is the condition for the modularity to choose the $C_2$-community. That means that if the $C_2$-community is approximately $t$ times larger than the $C_1$-community, the modularity function will produce bigger value for the case where the node $x$ is placed in the $C_1$-community, despite the fact that it has only 1 link with nodes from that community, compared to the $t$ links with nodes from the $C_2$ community. One can say that the modularity tends to make the communities equal.

\section{Computational complexity of the algorithm} \label{app:complexity}

The algorithm consists of 4 steps. We now analyze the running times of each step in order to determine the overall algorithm's complexity.

\subsection*{Influence matrix}

\begin{figure}[htbp]
  \begin{center}
  \includegraphics[scale=0.43]{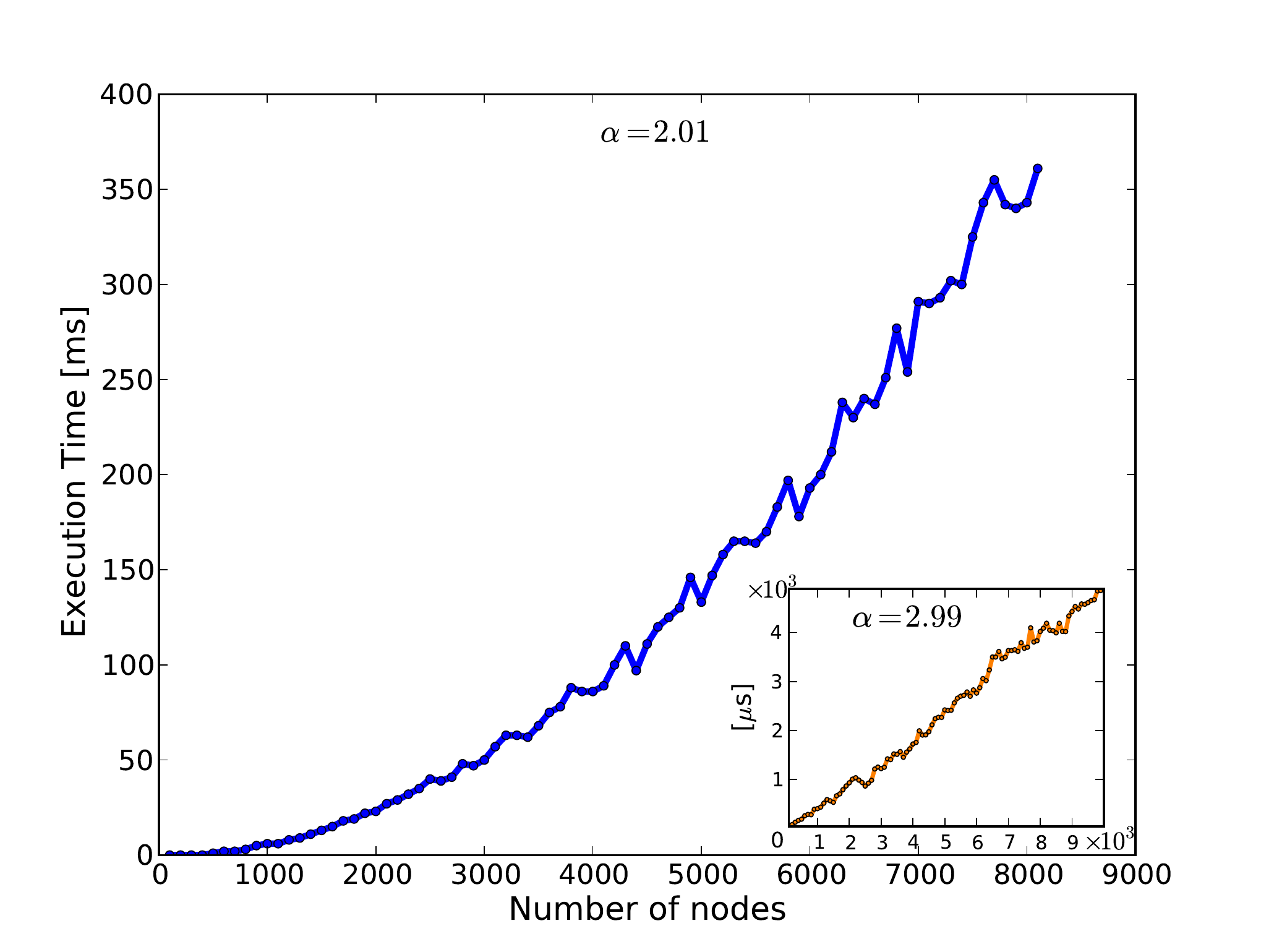}
  \end{center}
  \caption{(Color online) Execution time simulations for calculating the influence matrix. The inset shows that the running time grows linearly with the number of nodes when $\alpha=2.99$. On the other hand, the running time is $O(N^2)$ when $\alpha=2.01$.}
  \label{fig:step1}
\end{figure}

Building the weighted adjacency matrix $A$ is done by computing the number of mutual triangles or common neighbors between every pair of neighboring nodes in the network. Without loss of generality, we give a pseudo code for the undirected network case
\begin{algorithmic}
\FOR{each node $i$}
\FOR{each neighbor $j$ of $i$}
\STATE intersectNeighbors($i$,$j$)
\STATE \COMMENT{calculate $A_{ij}'$}
\ENDFOR
\ENDFOR
\end{algorithmic}
where $intersectNeighbors(i,j)$ finds the intersection between the set of neighbors of node $i$ and $j$ in time linear with the size of the sets, since we keep the sets sorted. Consequently, the running time is $\sum_i^Nk_i^2$, where $k_i$ is the degree of node $i$. Note that this is not the same as $N\cdot \left< k \right>^2$, where $\left< k \right>$ is the average node degree, since in real networks, the degree distribution is usually a power law distribution, $P(k)\sim k^{-\alpha}$, with the scaling factor $2<\alpha<3$. We have 
$$\sum_i^Nk_i^2=N\int k^2P(k)dk.$$
Since, this integral diverges, we have to approximate its lower and upper bound. We consider the network to be connected, meaning there exists a path between every pair of nodes in the network. That means the lowest degree in the network is 1, and that is the lower bound of the integral $k_{min}=1$. In \cite{newman-overview} an approximation $k_{max} \approx N^{\frac{1}{\alpha-1}}$ is derived. We take $P(k)=Ck^{-\alpha}$, where $C$ is a constant. Therefore, we have
\begin{eqnarray*}
\sum_i^Nk_i^2 &=& N\int_{k_{min}}^{k_{max}}k^2P(k)dk \\
&\approx & N\int_1^{N^{\frac{1}{\alpha-1}}}k^2P(k)dk = NC\int_1^{N^{\frac{1}{\alpha-1}}}k^{2-\alpha}dk \\
&=& NC\frac{k^{3-\alpha}}{3-\alpha}|^{N^{\frac{1}{\alpha-1}}}_1 \sim N\cdot N^{\frac{3-\alpha}{\alpha-1}}\\ 
&=& \left\{\begin{array}{l l}N^2 & \quad \text{if $\alpha$=2}\\N & \quad \text{if $\alpha$=3}\\\end{array}\right.
\end{eqnarray*}
Thus, the running time of computing the influence matrix varies from $O(N)$ to $O(N^2)$ depending on the scaling factor $\alpha$. This can be confirmed by Fig. \ref{fig:step1}, where we generated graphs with power-law degree distribution for $\alpha=2.01$ and $\alpha=2.99$. Each point is an average of 100 runs. As expected, the running time for $\alpha=2.01$ is quadratic and for $\alpha=2.99$ is linear.

\subsection*{Nodes' overall influences}

This process is actually a random walk process. If we have an undirected network, we even know the exact influences. So, the complexity is $O(c\cdot m)\approx O(m)\approx O(N)$. $c$ is the number of iterations until convergence, and its usually less than $50$ and $m\propto N$ in sparse graphs.

\subsection*{Leaders identification}

Here each node is in a battle with each of its potential parents, so clearly we have $O(N)$ complexity.

\subsection*{Computing the membership vectors}

\begin{figure}[htbp]
  \begin{center}
  \includegraphics[scale=0.43]{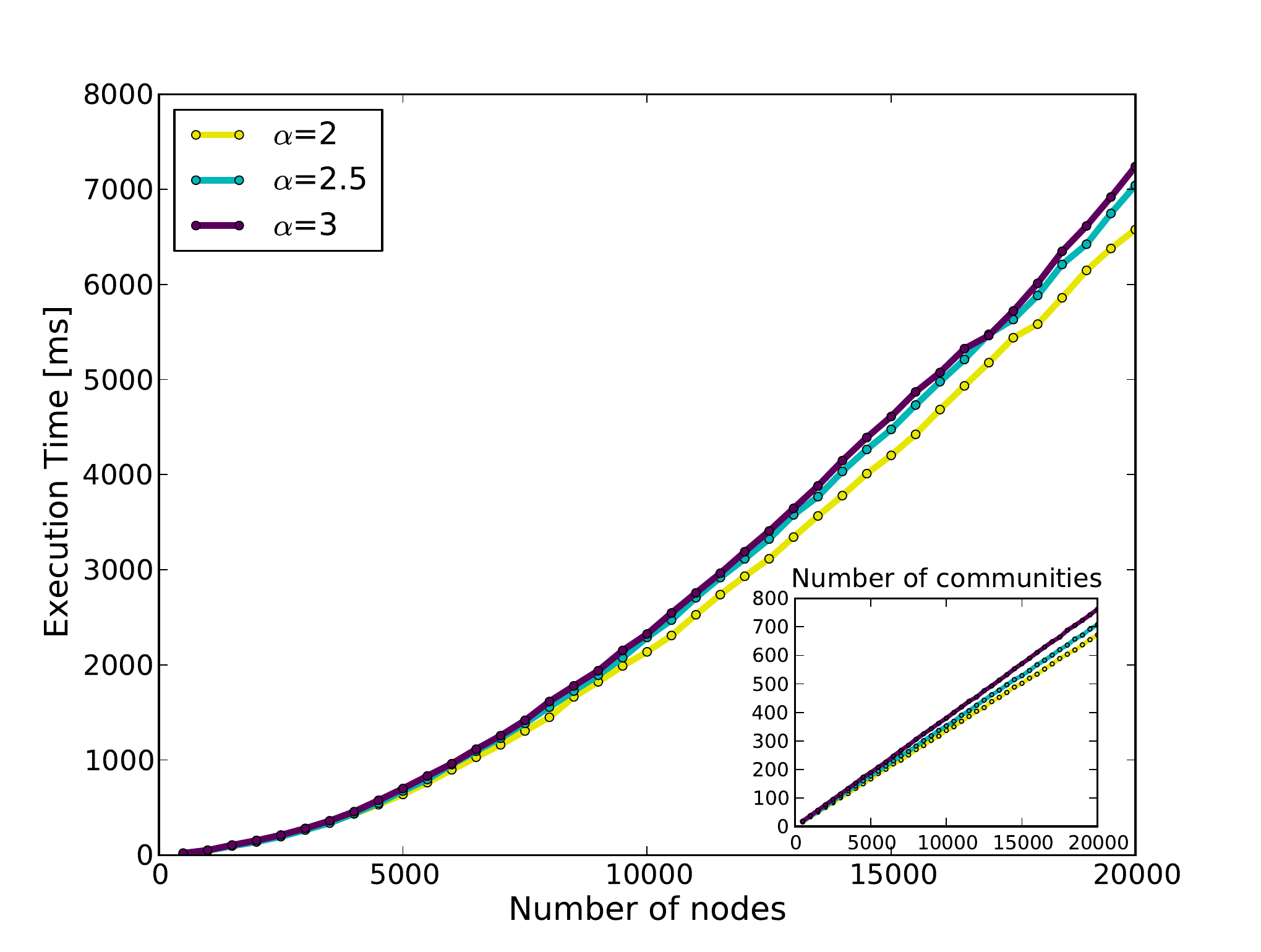}
  \end{center}
  \caption{(Color online) Running times of the algorithm on the LFR benchmark \cite{lfr}. The inset shows that the number of communities grows linearly with the number of nodes, because we restrict the community size. As a consequence, the complexity of the algorithm is $O(N^2)$.}
  \label{fig:lfr}
\end{figure}

This operation is very similar to the consensus linear process, with the difference of having a vector, instead of a single number, associated with each node. The complexity is $O(N\times L)$, where $L$ is the number of leaders. 
In Fig. \ref{fig:lfr} we show the execution time of this step on simulated LFR networks with power-law degree distribution for $\alpha=2, 2.5$ and $3$ \cite{lfr}. The parameters we use are similar to the ones in \cite{lfr}. Each point is an average of 100 runs. The average node degree is 20 and maximum degree is 50. The exponent of the power-law distribution of community size is 1, minimum community size is 20 and maximum community size is 100. The mixing parameter of every network is 0.3. Since we restrict the community size, the number of generated communities grows linearly with the number of nodes, i.e. $L\sim N$, thus, rendering the running time to quadratic. This comes only as a consequence of the application of the LFR benchmark and its parameters, and does not reflect any characteristics of our algorithm. In general the number of communities does not necessarily grow with the size of the network.

To conclude this section, the running times of the first and the last step are of the highest order, with execution times varying from $O(N)$ to $O(N^2)$, depending on the power-law exponent and the number of detected communities, respectively.  Thus, the overall complexity varies from $O(N)$ to $O(N^2)$ as well.

\end{document}